\documentstyle[mprocl]{article}

\def\Journal#1#2#3#4{{#1} {\bf #2}, #3 (#4)}

\def\be{\begin{equation}}
\def\ee{\end{equation}}
\def\bea{\begin{eqnarray}}
\def\eea{\end{eqnarray}}

\newcommand{\rea}{{\rm I\hspace{-0.5mm}R}}
\newcommand{\nat}{{\rm I\hspace{-0.4mm}l\hspace{-1.1mm}N}}

\newcommand{\dspa}[2]{\mbox{$(#1,{\cal #2})$}}
\newcommand{\fun}[3]{\mbox{$#1:#2  \rightarrow #3$}}

\begin{document}

\title{GRAVITATIONAL CONICAL BREMSSTRAHLUNG AND DIFFERENTIAL STRUCTURES}

\author{ J. GRUSZCZAK }

\address{Institute of Physics and Computer Science,
Cracow Pedagogical University, \\
ul. Podchora\.zych 2, 30-084 Cracow,  Poland}


\maketitle\abstracts{
Differential properties
of a spin 2 boson field $\psi_{\mu\nu}$ describing propagation of gravitational
perturbations
on a straight cosmic string's space-time background
are studied by means of methods of the differential spaces theory.
It is shown that this field is a smooth one in  the interior
of cosmic string's space-time and looses this property at the singular
boundary except for cosmic string space-times with the following
deficits of angle: $\Delta=2\pi (1-1/n) $, $n=1,2,\dots$.\\
A relationship between  smoothness of $\psi_{\mu\nu}$
at the singularity and
the gravitational conical bremsstrahlung effect is  discussed.
A physical interpretation of the smoothness notion is given.
It is also argued that the assumption
of smoothness of $\psi_{\mu\nu}$ at the singularity
plays  an equivalent role
to the Aliev and Gal'tsov "quantization" condition.\\
\\
To appear in the Proceedings of the 8th Marcel Grossman Meeting,
Jerusalem, June 1997 (World Scientific, Singapore).\\
\\
gr-qc/9801???}

\section{Gravitational Conical Bremsstrahlung Effect}  \label{intro}
Conical bremsstrahlung appears when a point mass is moving
in a space-time with the conical singularity.
Then the particle produces perturbations of
the background metric
propagating in form of  gravitational waves.
The effect was found by Aliev and Gal'tsov \cite{56af,43af} and
is an example of gravitational version
of the Aharonov-Bohm effect; radiative phenomena appear in absence of
local gravitational forces.
The total radiative energy emitted by
a particle moving near the straight cosmic string's conical singularity
vanishes  for the following special
values of the deficit of angle:
$\Delta =2\pi (1-1/n), n=1,2,... $.
The same effect holds for a K-G scalar and an electromagnetic fields.

\section{Differential Spaces and the Smoothness Notion}
Every space-time $M$ as a Lorentzian manifold is endowed with various
mathematical structures. Among other ones, there is a Sikorski's
 {\em differential structure} ${\cal C}$. Briefly speaking, it is a family
 of real functions on $M$ such that: a) ${\cal C}$ is closed with
 respect to
 localization, b) ${\cal C}$ is closed with respect to superposition
 with smooth functions on $\rea ^n$ for any $n\in\nat$ and
 c) $M$ is equipped with topology such that every function from ${\cal C}$
 is continuous \cite{12af}.

The pair \dspa{M}{C} \ is said to be a {\em differential space} and is
a notion more general than the manifold one; every manifold is a d-space
but not every d-space is a manifold. For example, the space-time of
straight cosmic string is a d-space which is a manifold in contrast to
the cosmic string's space-time with the conical singularity which is not a
manifold but is still a d-space.

In the theory of d-spaces the smoothness notion plays the fundamental
role. It is a generalization of the well known one from the manifold
theory.
Thus,
a real function \fun{\phi}{M}{\rea} \ is said to be a smooth one
on $(M, {\cal C})$
if $\phi\in {\cal C}$ while
a vector  field $\fun{{\bf V}}{\cal C}{\cal C}$ \
is said to be smooth one on $(M, {\cal C})$
if ${\bf V}({\cal C})\subset {\cal C}$.

The smoothness definition for tensor fields is somewhat more
complicated \cite{12af,18af}.
But, for the tensor potential $ \psi_{\mu\nu}$ important in this paper
the test of smoothness is based (generally speaking) on verification of
the following condition:
$$\psi_{\mu\nu}V^\mu V^\nu\in {\cal C},$$
for every smooth vector field $\fun{\bf V}{{\cal C}}{{\cal C}}$.
In the  complex  case, a complex tensor field is a smooth one if its real
and imaginary parts  are smooth real tensor fields.

\section{Smoothness of Gravitational Perturbations}
Gravitational perturbations  of the straight cosmic string's space-time
produced by a moving particle are described by the potential
$\psi_{\mu\nu}$ satisfying in the De-Donder gauge
$\nabla_{_\mu} \psi^{_{\mu\nu}}=0$
the following field equation:
$\nabla_{_\mu}\nabla^{_\mu}\psi_{_{\rho\sigma}}=0$.
The elementary solutions of the wave equation are denoted by
$\psi_{\mu\nu}^{{(\epsilon  \beta  l)}} $.

\begin{itemize}
\item[]{\bf Proposition:}
{\em
The elementary solutions $\psi_{\mu\nu}^{{(\epsilon  \beta  l)}} $
are smooth complex tensor fields on the  cosmic string
space-time manifold.
}
\end{itemize}

Thus, the elementary solutions have the smoothness
property characteristic for the theory of differential spaces.
The space time of cosmic string with singularity is not a manifold
but is still a d-space.
Can the smoothness property of
$\psi_{\mu\nu}^{{(\epsilon  \beta  l)}}$ survive after prolongation
to singularity?

\begin{itemize}
\item[]{\bf Theorem:}
{\em
The elementary solutions $\psi_{\mu\nu}^{{(\epsilon  \beta  l)}} $
after prolongation to singular boundary
are smooth complex tensor fields on the  differential space of cosmic string
with singularity only for the following deficits of angle
$\Delta =2\pi (1-1/n), n=1,2,... $
For the remaining deficits of angle they are not smooth complex
tensor fields.
}
\end{itemize}
The proof is based on verification of the condition mentioned in the
previous section.
The similar theorem holds for a scalar and an electromagnetic fields
\cite{63af,69af}.

\section{Discussion}

The family of all straight cosmic string's space-times can be divided
into two separate subfamilies. For the first one
($\Delta =2\pi (1-1/n), n=1,2,... $) every elementary perturbation
$\psi_{\mu\nu}^{{(\epsilon  \beta  l)}} $ is
a smooth complex tensor field on the
whole cosmic string's space-time background including singularity.
For the second one
($\Delta \ne 2\pi (1-1/n), n=1,2,... $)
not every elementary perturbation has the  smoothness property.
According to the results of Aliev and Gal'tsov mentioned in the
Section \ref{intro},
this division is also natural  from strictly physical view
point. Namely, the conical bremsstrahlung effect existing for the second
subfamily vanishes for the first one. The mathematical
reason of this coincidence are differential properties of the elementary
solutions at singularity (smoothness). They are mirrored in the Aliev
and Gal'tsov results through the radiative Green function constructed with
help of $\psi_{\mu\nu}^{{(\epsilon  \beta  l)}} $.

The above coincidence enable the following
physical interpretation of the smoothness notion. Every massive body
moving in the gravitational field of cosmic string emits, in general,
gravitational waves. One can expect that such situation will continue
until the deficit of angle reaches one of the distinguished values:
$\Delta = 2\pi (1-1/n), n=1,2,... $.
Then, the system (a particle moving in gravitational field of cosmic string)
becomes stationary; the conical bremsstrahlung effect vanishes.
One can express the same
in terms of strictly geometrical notions
from the theory of d-spaces.  The
system will radiate till every non smooth elementary mode becomes smooth
one.
The nonsmoothness of the elementary solutions indicates that the
system is not stationary.
The system becomes stationary
 if every elementary
solution $\psi_{\mu\nu}^{{(\epsilon  \beta  l)}} $ is a smooth tensor
field on the whole space-time of cosmic string with singularity.

It is also possible an another much radical interpretation.
The smoothness of tensor fields is a natural requirement within the
theory of differential spaces. In some sense the non smoothness is a
symptom of the field theory consistency breaking.
The theory of gravitational perturbations of the straight
cosmic string's metrics with regularity conditions at the singularity
\cite{56af,43af,68af} can be treated as a field theory on a d-space which is
not a manifold.
Therefore, it is natural to assume that the
gravitational perturbations
have to be smooth one on the whole cosmic string's space-time with
singularity. Such assumption is equivalent to the Aliev and Gal'tsov
condition of vanishing the conical bremsstrahlung effect and leads to
the following discrete spectrum of the deficit of angle:
$\Delta =2\pi (1-1/n), n=1,2,3,...$.

\section*{Acknowledgments}
I thank Prof. K.Ruebenbauer for helpful discussion.
This work was supported by the KBN Research Project no. 2 P03D 02210.


\section*{References}

\begin{thebibliography}{1}

\bibitem{56af}
A.N. Aliev and D.V. Gal'tsov, \Journal{\em Ann. Phys.}{193}{142}{1989}.

\bibitem{43af}
A.N. Aliev, \Journal{\em Class. Quantum Grav.}{10}{2531}{1993}.

\bibitem{12af}
J. Gruszczak, M. Heller and P. Multarzynski,
\Journal{\em J. Math. Phys.}{29}{2576}{1988}.

\bibitem{18af}
M. Heller, P. Multarzynski and W. Sasin,
\Journal{\em Acta Cosmologica}{16}{53}{1989}.

\bibitem{63af}
J. Gruszczak, \Journal{\em Acta Cosmologica}{21{\rm (2)}}{219}{1995}.

\bibitem{69af}
J. Gruszczak, {\em gr-qc/9609070}.

\bibitem{68af}
B.S. Kay and U.M. Studer,
\Journal{\em Commun. Math. Phys.}{139}{103}{1991}.

\end{thebibliography}
\end{document}